\documentclass[12pt,preprint]{aastex}
\usepackage{amsmath}
\usepackage{graphicx}
\usepackage{float}

\begin{document}
\newcommand{\vv}{\textrm{v}}
\newcommand{\sname}{OGLE-LMC-ECL-11893}
\newcommand{\sshort}{11893}
\title{OGLE-LMC-ECL-11893: The discovery of a long-period eclipsing binary with a circumstellar disk} 
\author{Subo Dong\altaffilmark{1}, Boaz Katz\altaffilmark{2,3}, Jose L. Prieto\altaffilmark{4,5}, Andrzej Udalski\altaffilmark{6,7}, Szymon Kozlowski\altaffilmark{6,7}, R. A. Street\altaffilmark{8,15}, D. M. Bramich\altaffilmark{9,15}, Y. Tsapras\altaffilmark{8,10,15}, M. Hundertmark\altaffilmark{11,15},  C. Snodgrass\altaffilmark{12,15}, K. Horne\altaffilmark{11,15}, M. Dominik\altaffilmark{11,13,15}, R. Figuera Jaimes\altaffilmark{14,11,15}}
\altaffiltext{1}{Kavli Institute for Astronomy and Astrophysics, Peking University, Yi He Yuan Road 5, Hai Dian District, Beijing 100871, China}
\altaffiltext{2}{Institute for Advanced Study, 1 Einstein Dr.}
\altaffiltext{3}{John N. Bahcall Fellow}
\altaffiltext{4}{Department of Astrophysical Sciences, Princeton University, 4 Ivy Lane, Peyton Hall, Princeton, NJ 08544}
\altaffiltext{5}{Carnegie-Princeton Fellow}
\altaffiltext{6}{Warsaw University Observatory, Al. Ujazdowskie 4, 00-478 Warszawa, Poland}
\altaffiltext{7}{The OGLE Collaboration}
\altaffiltext{8}{Las Cumbres Observatory Global Telescope Network, 6740 Cortona Drive, suite 102, Goleta, CA 93117, USA}
\altaffiltext{9}{Qatar Environment and Energy Research Institute, Qatar Foundation, Tornado Tower, Floor 19, P.O. Box 5825, Doha, Qatar}
\altaffiltext{10}{School of Physics and Astronomy, Queen Mary University of London, Mile End Road, London E1 4NS, UK}
\altaffiltext{11}{SUPA, School of Physics \& Astronomy, University of St Andrews, North Haugh, St Andrews KY16 9SS, UK}
\altaffiltext{12}{Max Planck Institute for Solar System Research, Justus-von-Liebig-Weg 3, 37077 G\"ottingen, Germany}
\altaffiltext{13}{Royal Society University Research Fellow}
\altaffiltext{14}{European Southern Observatory, Karl-Schwarzschild-Stra{\ss}e 2, 85748~Garching~bei~M\"unchen, Germany}
\altaffiltext{15}{The RoboNet Collaboration}

\begin{abstract}
We report the serendipitous discovery of a disk-eclipse system \sname. 
The eclipse occurs with a period of 468 days, a duration of about 15 days and a deep (up to $\Delta m_I\approx 1.5$), 
peculiar and asymmetric profile. A possible origin of such an eclipse profile involves a circumstellar disk. The presence of the disk is confirmed by the H-$\alpha$ line profile from the follow-up spectroscopic observations, and the star is identified as Be/Ae type.
Unlike the previously known disk-eclipse candidates, 
the eclipses of \sname\,\, retain the same shape throughout the span of $\sim 17$ years (13 orbital periods), 
indicating no measurable orbital precession of the disk.
\end{abstract}

\section{Identification of the peculiar eclipses of \sname}
\begin{figure}
\centering
\includegraphics*[width=1.0\textwidth]{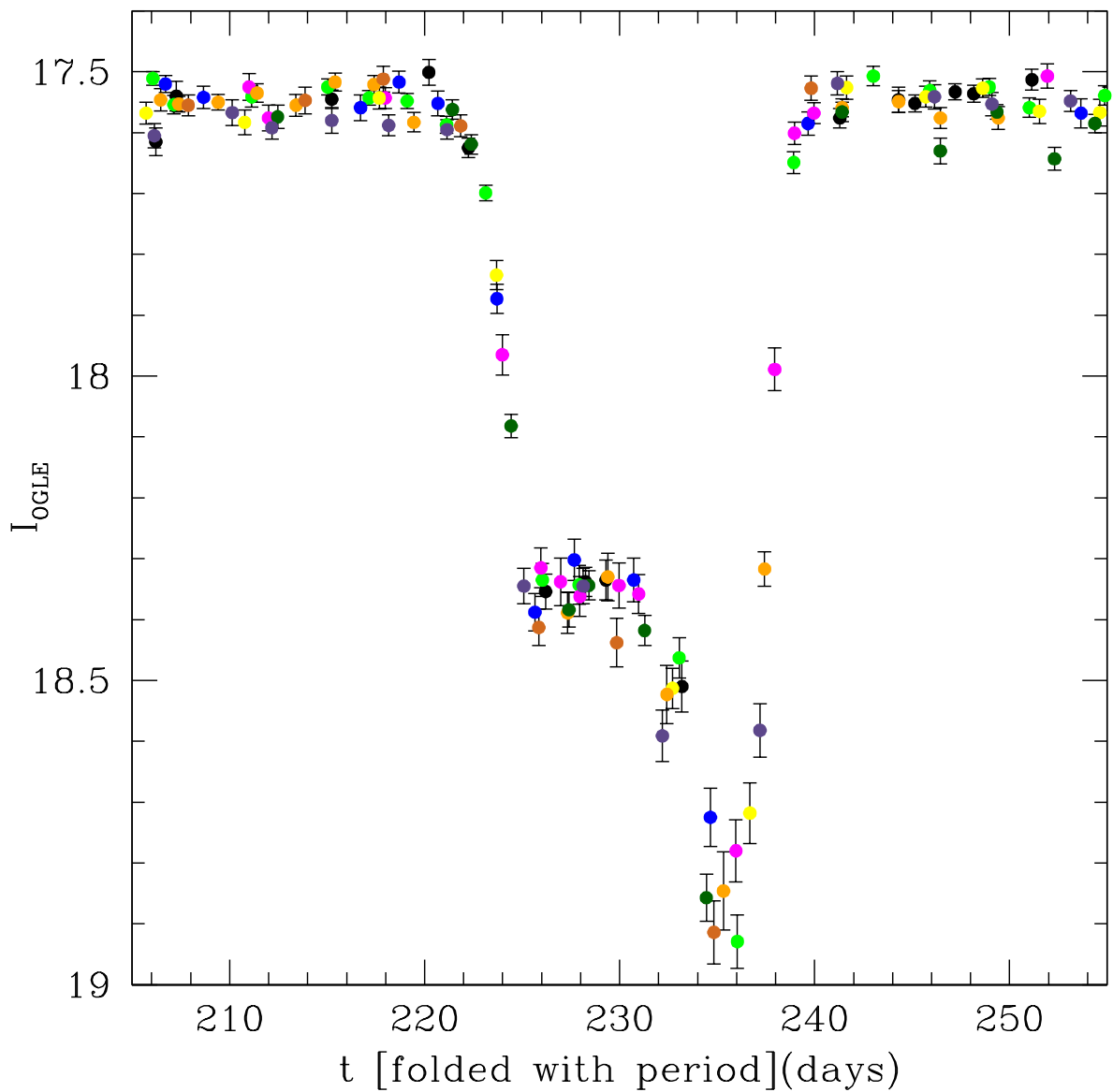}
\caption{Period-folded OGLE $I$-band light curve of \sname\,\, with respect to 
${\rm HJD}_0 = 2454245.5663$. Photometry from different orbital periods are 
shown in different colors. The observations by OGLE-II and OGLE-III span 
$\sim 12$ years and contain 9 eclipse occurrences. The peculiar and asymmetric
eclipse profile is unchanged within measurement uncertainties during the 
span of observations.
\label{fig:ogle}}
\end{figure}

The OGLE LMC eclipsing binary (EB) catalogue \citep{Graczyk11} reported the discovery of 26121 EBs toward the LMC. 
In November 2011, while searching the catalogue for long-period (period $P>100$ days), highly-eccentric binaries \citep{Dong13}, we 
serendipitously identified a very unusual eclipsing binary \sname\,\, (RA: 05:17:21.17 DEC: -69:00:55.7; J2000) by visual inspection. The period-folded I-band light curve shows an eclipse with a peculiar shape  (Figure. \ref{fig:ogle}). The data are taken from phases II and III of the OGLE survey \citep{ogleii, ogleiii} and span 12 years (further data from OGLE-IV survey are discussed in Sec. 2). The period is 468.124 days, and the baseline $I$-band magnitude is about 17.6. Each eclipse occurrence is shown with a different color. The eclipse bears some close similarity with EE Cephei, an eclipsing Be star with a circumstellar disk \citep{eecephei} but also has some unique characteristics. In the following, we list several striking features of the \sname\,\, eclipses:
\begin{enumerate}
\item The eclipse profile, which spans about 15 days, has an unusual asymmetric shape that cannot be produced by a spherical occulting star. The shape of the profile bears close similarity with EE Cephei during its 2008/9 eclipse (e.g., see the right panel of Fig. 3 in \citealt{eecephei}) and the duration of EE Cephei is similar too ($\sim 30 {\,\rm days}$). The ingress and egress of \sname\,\, are 
steeper than EE Cephei.
\item The eclipse is deep, with a maximal depth of $\sim 1.5$ magnitudes or equivalently, $75\%$ of the I-band light is occulted. This is comparable to that of EE Cephei, whose eclipse depths vary between $\sim 0.5 - 2.0$ mag \citep{eecephei}. 
\item The shape of the eclipse does not change with time. The 9 eclipses spanning 17 years detected so far have exactly the same profile within measurement uncertainties. This is in contrast with EE Cephei and three other well-known disk-eclipse events ($\varepsilon$ Aurigae, OGLE-LMC-ECL-17782, KH 15D), whose eclipse depths vary from one orbital period to another. These variations are generally attributed to the precessions of the disks.   
\end{enumerate}  
The similarities between \sname\,\,and EE Cephei hint that they have similar physical origins, possibly disk-eclipsing events, as suggested by \citet{eecephei} for EE Cephei.

\section{Follow-up observations}
\begin{figure}
\centering
\includegraphics[width=0.5\textwidth]{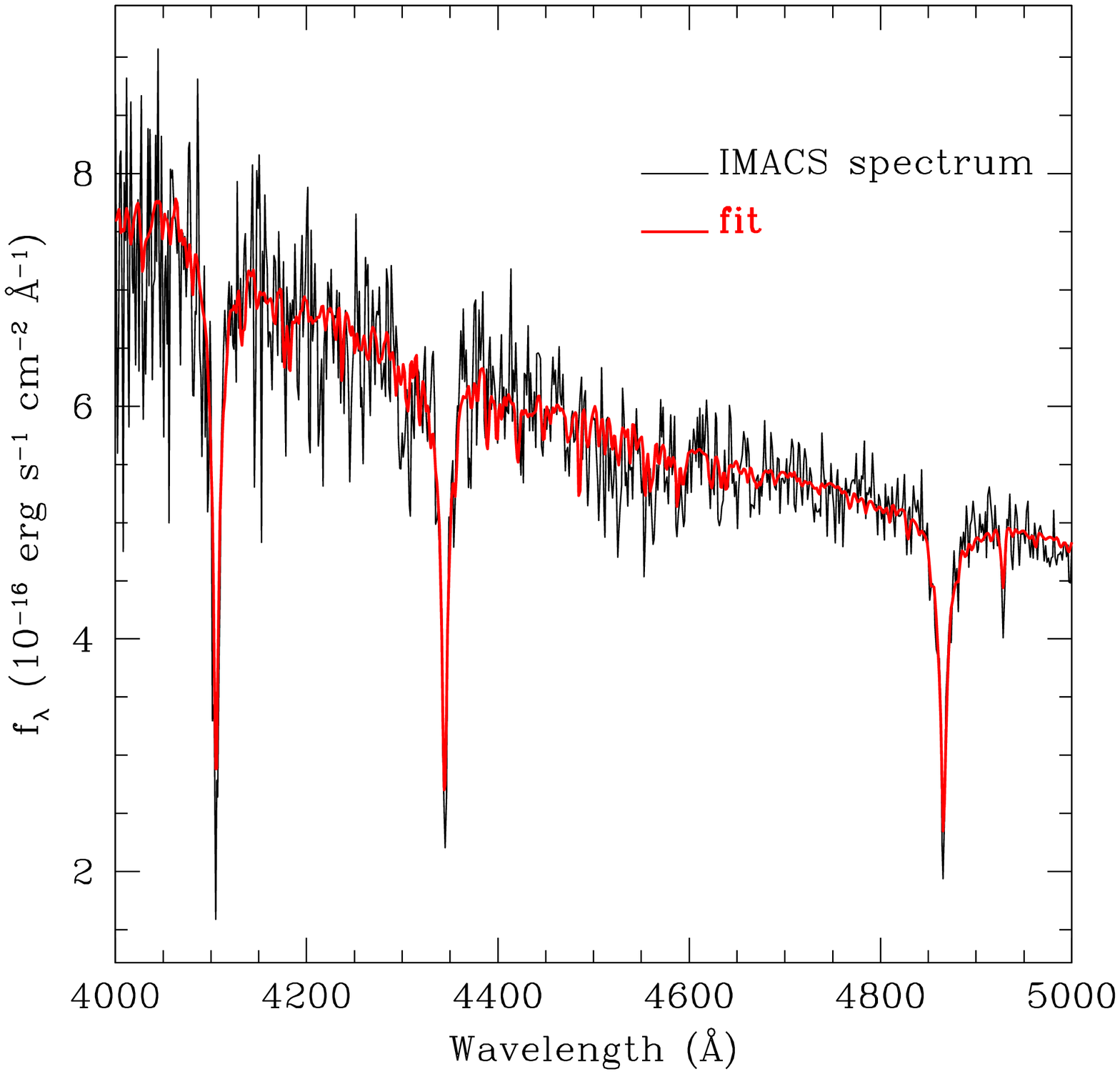}
\includegraphics[width=0.5\textwidth]{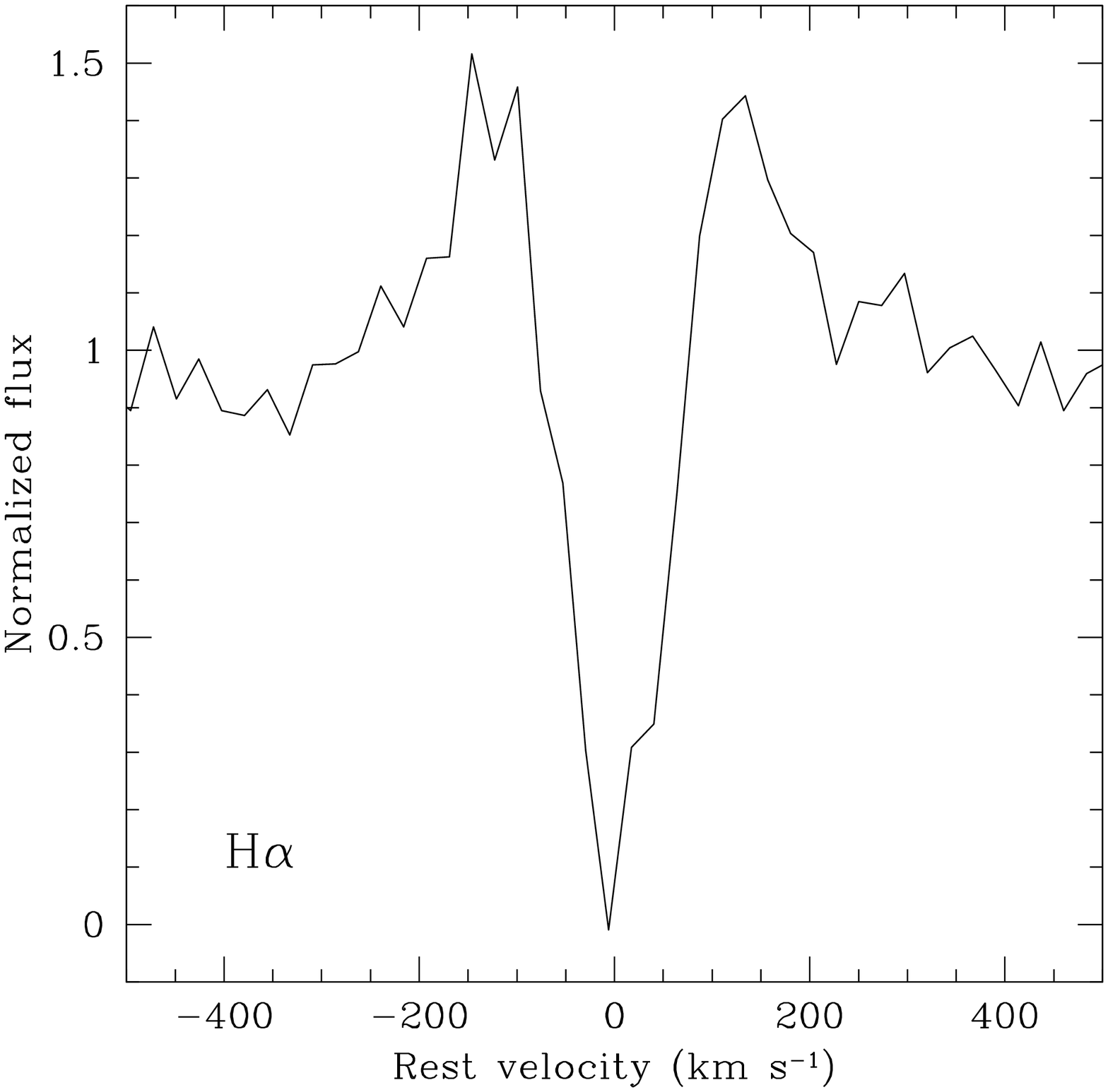}
\caption{Upper Panel: Spectrum of \sname\,\,taken by IMACS (black). The best-fit model from Ulyss (red) suggests it is a late B/early A-type giant. Lower Panel: H-$\alpha$ line profile in the MagE spectrum that is characteristic of Ae/Be stars confirms the existence of a circumstellar disk \citep{halpha}. The same profile also clearly shows up in the IMACS spectrum though the H-$\alpha$ region is near the edge of the CCD. The H-$\alpha$ profile is typical for an edge-on Be star, similar to that of $\varepsilon$ Aurigae (see bottom left panel of Fig. 8 on page 241 in \citealt{halpha}), for which the disk was directly confirmed by infrared interferometer imaging \citep{Kloppenborg10}. 
\label{fig:spectrum}}
\end{figure}

We obtained two optical spectra of OGLE-LMC-ECL-11893, first on 2011
Dec. 23 with IMACS f/2 (Dressler et al. 2011) mounted on the Magellan
Baade 6.5m and the second on 2013 Jan. 7 with the Magellan Echellette
spectrograph (MagE; Marshall et al. 2008).  For IMACS f/2 spectrum we
used the 300 l/mm grism and 0.7\arcsec slit, which gives resolution R = 1800
and wavelength coverage $3700-9500$~\AA. We show the IMACS spectrum of
the star in the upper panel of Figure \ref{fig:spectrum}.  MagE with 1\arcsec slit gives $R=4800$ 
and wavelength coverage $3300-10000$~\AA. The IMACS spectrum was 
reduced with standard techniques in IRAF. The MagE spectrum was reduced with the Carnegie pipeline from D. Kelson. 

Using the Ulyss package which constrains stellar parameters by fitting stellar spectra with the Elodie spectroscopy library \citep{Koleva09}, we find that the spectrum is consistent with that of a late B/early A-type giant with $T_{\rm eff}\sim 10000 K$. The Radial Velocity (RV) of the star measured from the MagE spectrum is $270$~km/s, consistent with the RV of the LMC. In both spectra, we find H-$\alpha$ line profiles characteristic of circumstellar disks \citep{halpha}, confirming the existence of the disk indicated by the eclipse, and the system is of Be/Ae type. The H-$\alpha$ profile in the MagE spectrum is shown in the lower panel of Figure \ref{fig:spectrum}. The H-$\alpha$ profile is similar to that of $\varepsilon$ Aurigae \citep{halpha}, typical for Be star observed edge-on (see Fig. 3 in \citealt{review}). $\varepsilon$ Aurigae is a disk-eclipsing event for which the disk was directly confirmed by infrared interferometry \citep{Kloppenborg10}. 

\begin{figure}
\centering
\includegraphics[width=1.0\textwidth]{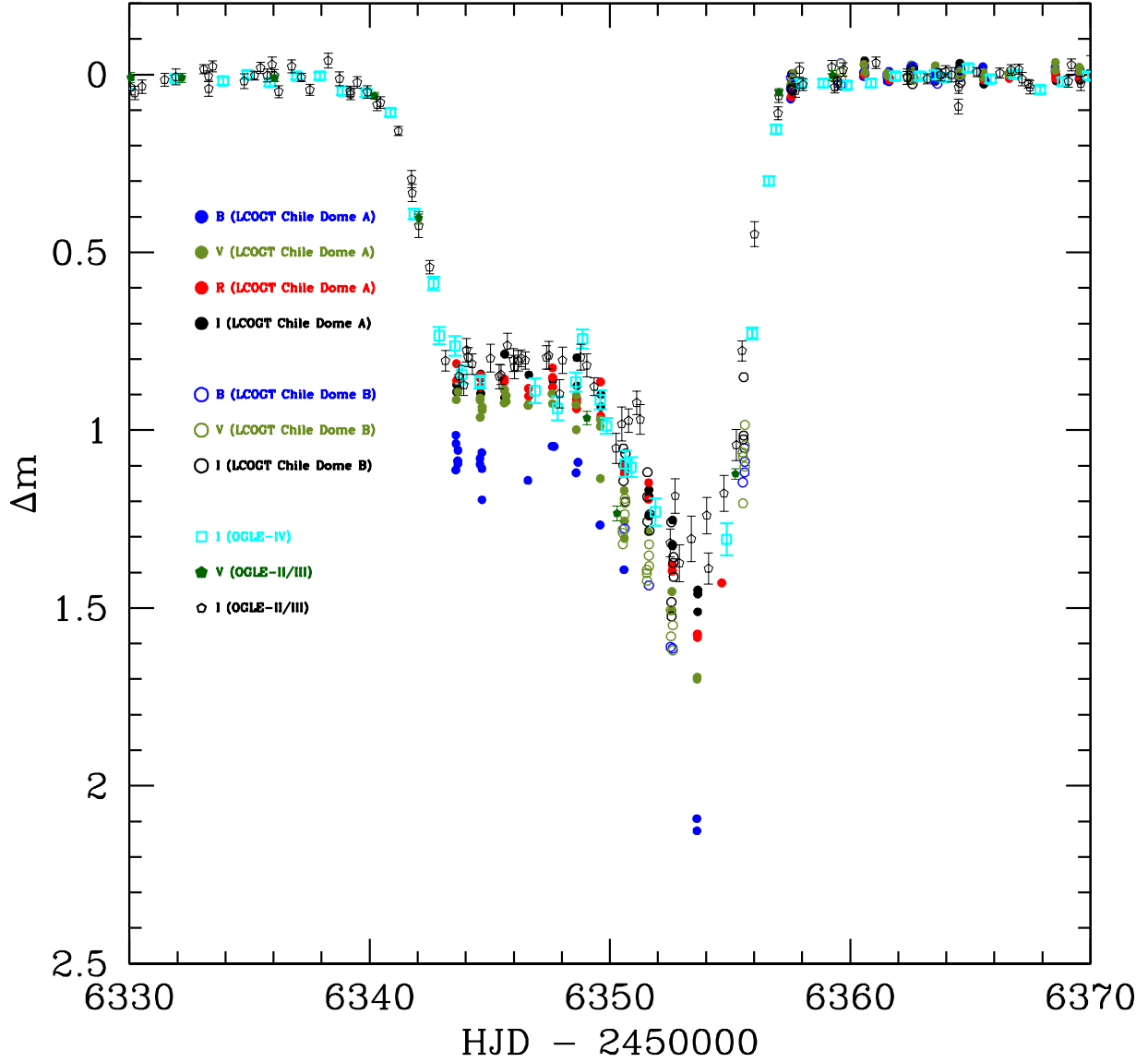}
\caption{Multi-band follow-up light curves of the \sname\,\, $BVRI$ filters 
(in blue, olive-green, red, black) were taken by LCOGT Chilean 1-m 
telescopes (Dome A: filled circles, Dome B: open circles) during the 2013 eclipse.
The OGLE-IV $I$-band observations (cyan open squares) were taken in the 2013 
and 2011 seasons. The OGLE-III and OGLE-II $I$-band (black open pentagons) 
and $V$-band (dark green filled pentagons) observations are also shown for 
comparison. Observations which were not taken in 2013 are shifted by periods 
to match the time of the eclipse in 2013.
\label{fig:Multiband}}
\end{figure}

Multi-band photometry obtained using the 1m telescopes (Domes A and B) at
LCOGT's site in Chile (CTIO) during February and March 2013 (at the early commissioning stages of the telescopes) are shown in
Figure \ref{fig:Multiband} and given in Table 1. The observations were conducted through the
Observation Control interface developed for the RoboNet project, taken
in semi-robotic mode as the underlying software was in the process of
being upgraded to integrate with the new 1m network at the time. The
image data were preprocessed using LCOGT's Standard Pipeline, which is
based around ORAC-DR, after which the RoboNet pipeline (based on DanDIA by \citealt{dandia})
was used for difference image analysis (DIA) photometry.

The different bands (filled and empty dots) are compared to the folded $I$ and $V$ bands OGLE II, III and IV photometry
 (empty squares and filled
and empty pentagons). Given the photometric uncertainties in the LCOGT data, there is reasonable 
agreement between the LCOGT data and the
OGLE data for the $V$- and $I$- bands where they could be directly compared.
While most bands have a roughly similar profile, the eclipse depth seems to increase for bluer bands (from $I$, $R$, $V$ to $B$). In particular, the $B$-band seems to
have a consistently deeper eclipse, and the two $B$-band measurements during
HJD 2456353, in the deepest point of the eclipse seem to be particularly deep
compared to other bands. The colour variations during the eclipse events may be related to several physical properties of the system such as possible temperature variations across the eclipsed areas, the opaqueness of the disk as a function of wavelength, the limb darken profile of the eclipsed stars, etc.
Further high-precision observations are
required to accurately measure these differences and study their
implications.

The next two eclipses of OGLE-LMC-ECL-11893 will be around HJD $\sim 
2456818$ (June 2014) and $\sim 2457286$ (September 2015). The former one,
however, will be practically unobservable as it will be too close to the Sun. We encourage intense multi-band photometric and spectroscopic monitoring during the eclipse. In addition, long-term RV monitoring will be useful to determine the orbital parameters and the mass of the eclipsing component.

\section{Discussion}
Assuming a circular orbit, the size of an eclipsing region is of the order:
$D\sim 2\pi a  \Delta{t}  /P= \Delta{t} (2\pi GM_{\rm tot}/P)^{1/3}\sim0.35 (M_{\rm tot}/3M_{\odot})^{1/3}{\rm AU}$ for eclipse duration $\Delta{t} \sim 15d$ and $P=468d$, where $M_{\rm tot}$ is the total mass of the system and $a$ is the semi-major axis.

Three well known periodic disk-eclipse EB candidates are $\varepsilon$ Aurigae 
($P=27\,\rm yr$), EE Cephei ($P=5.6\,\rm yr$), and OGLE-LMC-ECL-17782 ($P=13.4\,\rm day$, identified by \citealt{Graczyk11} in the OGLE LMC EB catalogue). 
 The remarkably complicated eclipses of KH 15D are interpreted as a circumbinary disk eclipsing a pair of stars \citep{kh15d}. Evidences for disk-eclipsing nature of YSOs YLW 16A and WL 4 have been presented in \citet{plavchan13} and \citet{parks14}.  An additional disk-eclipse candidate with one occurrence of eclipse was suggested by \citet{Mamajek12}. A LMC Be star FTS ID 78.5979.72 which had a sudden transition from no measurable variabilities to eclipsing with short time-variabilities was reported in \citet{macho}. 

The period, depth, eclipse duration and stellar type of \sname\,\, are similar to those of EE Cephei suggesting a similar origin for the peculiar eclipses in both systems. The H-$\alpha$ line profile is similar to that of $\varepsilon$ Aurigae. One important difference is that the shape of \sname\,\,appears to be constant while those of EE Cephei, $\varepsilon$ Aurigae, OGLE-LMC-ECL-17782 and KH 15D change between eclipses, interpreted as precessions of disks. \sname\,\, is exceptional in this respect that it does not have measurable precession. Another interesting qualitative difference between \sname\,\, and EE Cephei is that the former has relatively short durations of the ingress and egress than the latter.
The disk-eclipse scenario along with other physical properties of the system will be studied in depth by \citet{scott2014}. 


The serendipitous identification of \sname\,\, demonstrates the power of the OGLE survey in making discoveries of unanticipated variables, thanks to its long-term, high-cadence and high quality photometry of a large number of stars. This  has implications for planning future large time-domain surveys such as LSST because the overwhelming majority of such systems in the Galaxy would lie in the inner two quadrants of the Galactic
disk, a region that was excluded in the recently published LSST planning (See Figure 1 of \citealt{gould} and Figure 4.4 of \citealt{lsst}).

\acknowledgments   

We thank Cullen Blake for stimulating discussion on alternative models and the participants of a lively IAS astro-coffee session held in Dec 2011. We are grateful to Alice Quillen and Josh Winn for helpful comments. S.D. is supported by ``the Strategic Priority Research Program-The Emergence of Cosmological Structures'' of the Chinese Academy of Sciences (Grant No. XDB09000000). The OGLE project has received funding from the European Research Council under the European Community's Seventh Framework Programme (FP7/20072013)/ERC grant agreement No. 246678 to A.U. MD, RFJ, KH, RAS, CS, and YT are supported by NPRP grant NPRP-09-476-1-78 from the Qatar National Research Fund (a member of Qatar Foundation). DMB's contribution to this publication was made possible by NPRP grant \# X-019-1-006 from the Qatar National Research Fund (a member of Qatar Foundation). CS has received funding from the European Union Seventh Framework Programme (FP7/2007-2013) under grant agreement no. 268421.

\bibliographystyle{apj}

\begin{thebibliography}{21}

\bibitem[Abate et al.(2012)]{lsst} Abate, A., Aldering, G., Allen, S.W., et al. 2012, arXiv:1211.0310

\bibitem[Bramich (2008)]{dandia} Bramich D. M., 2008, MNRAS, 386, L77

\bibitem[Dong et al.(2013)]{Dong13} Dong, S., Katz, B., 
\& Socrates, A.\ 2013, \apjl, 763, L2 

\bibitem[Dressler et al.(2011)]{Dressler11} Dressler, A., Bigelow, 
B., Hare, T., et al.\ 2011, \pasp, 123, 288 

\bibitem[Ga{\l}an et 
al.(2012)]{eecephei} Ga{\l}an, C., Miko{\l}ajewski, M., Tomov, T., et al.\ 2012, \aap, 544, A53 

\bibitem[Gould(2013)]{gould} Gould, A.\ 2013, arXiv:1304.3455 

\bibitem[Graczyk et al.(2011)]{Graczyk11} Graczyk, D., 
Soszy{\'n}ski, I., Poleski, R., et al.\ 2011, \actaa, 61, 103 

\bibitem[Kloppenborg et al.(2010)]{Kloppenborg10} Kloppenborg, B., 
Stencel, R., Monnier, J.~D., et al.\ 2010, \nat, 464, 870  

\bibitem[Koleva et 
al.(2009)]{Koleva09} Koleva, M., Prugniel, P., Bouchard, A., \& Wu, Y.\ 2009, \aap, 501, 1269 

\bibitem[Mamajek et al.(2012)]{Mamajek12} Mamajek, E.~E., 
Quillen, A.~C., Pecaut, M.~J., et al.\ 2012, \aj, 143, 72 

\bibitem[Marshall et al.(2008)]{Marshall08} Marshall, J.~L., 
Burles, S., Thompson, I.~B., et al.\ 2008, \procspie, 7014,  

\bibitem[Parks et al.(2014)]{parks14} Parks, J.~R., Plavchan, 
P., White, R.~J., \& Gee, A.~H.\ 2014, \apjs, 211, 3

\bibitem[Plavchan et 
al.(2013)]{plavchan13} Plavchan, P., G{\"u}th, T., Laohakunakorn, N., \& Parks, J.~R.\ 2013, \aap, 554, A110

\bibitem[Silaj et al.(2010)]{halpha} Silaj, J., Jones, C.~E., 
Tycner, C., Sigut, T.~A.~A., \& Smith, A.~D.\ 2010, \apjs, 187, 228

\bibitem[Slettebak(1979)]{review} Slettebak, A.\ 1979, \ssr, 
23, 541

\bibitem[Struble et al.(2006)]{macho} Struble, M.~F., 
Galatola, A., Faccioli, L., Alcock, C., \& Cruz, K.\ 2006, \aj, 131, 2196

\bibitem[Udalski et al.(1997)]{ogleii} Udalski, A., Kubiak, 
M., \& Szymanski, M.\ 1997, \actaa, 47, 319 

\bibitem[Udalski et al.(2008)]{ogleiii} Udalski, A., Szymanski, 
M.~K., Soszynski, I., \& Poleski, R.\ 2008, \actaa, 58, 69

\bibitem[Winn et al.(2006)]{kh15d} Winn, J.~N., Hamilton, 
C.~M., Herbst, W.~J., et al.\ 2006, \apj, 644, 510

\end{thebibliography}

\clearpage
\begin{table*}
\caption{Time-series $B$, $V$, $R$ and $I$ photometry for OGLE-LMC-ECL-11893 from the LCOGT network of telescopes. The instrumental $m_{\mbox{\scriptsize ins}}$ magnitudes are listed in column 4 corresponding to the LCOGT site/dome/telescope/instrument code, filter, and epoch of mid-exposure listed in columns 1-3, respectively. The uncertainty on $m_{\mbox{\scriptsize ins}}$ is listed in column 5. For completeness, we also list the quantities $f_{\mbox{\scriptsize ref}}$, $f_{\mbox{\scriptsize diff}}$ and $p$ in columns 6, 8 and 10, along with the uncertainties $\sigma_{\mbox{\scriptsize ref}}$ and $\sigma_{\mbox{\scriptsize diff}}$ in columns 7 and 9. Note that $m_{\mbox{\scriptsize ins}}(t) = 25 - 2.5 \log ( f_{\mbox{\scriptsize ref}} + \frac{f_{\mbox{\scriptsize diff}}(t)}{p(t)} )$ where $t$ is time. Note that these magnitudes are not calibrated to standard systems and they should be used only for studying the relative flux changes. The calibrated magnitudes of the system at the baseline are: $I = 17.55 \pm 0.03$, $V=17.74\pm0.02$ and $B=17.95\pm0.04$. This is an extract from the full table, which is available with the electronic version of the article.}
\centering

\begin{tabular}{cccccccccc}
\hline
Telescope & Filter & HJD & $m_{\mbox{\scriptsize ins}}$ & $\sigma_{m}$ & $f_{\mbox{\scriptsize ref}}$ & $\sigma_{\mbox{\scriptsize ref}}$ & $f_{\mbox{\scriptsize diff}}$ & $\sigma_{\mbox{\scriptsize diff}}$ & $p$ \\
ID        &        & (d) & (mag)                        & (mag)        & (ADU s$^{-1}$)               & (ADU s$^{-1}$)                    & (ADU s$^{-1}$)                & (ADU s$^{-1}$)                     &     \\
\hline
lsc-doma-1m0-05-kb78 & B & 2456343.58641 & 21.502 & 0.034 & 69.323 & 1.879 & $-$43.868 & 0.773 & 0.9913 \\
lsc-doma-1m0-05-kb78 & B & 2456343.59010 & 21.404 & 0.032 & 69.323 & 1.879 & $-$41.498 & 0.804 & 0.9908 \\
\vdots   & \vdots & \vdots & \vdots & \vdots & \vdots   & \vdots & \vdots   & \vdots & \vdots \\
lsc-doma-1m0-05-kb78 & V & 2456343.60545 & 21.035 & 0.021 & 38.004 & 3.224 & 0.599 & 0.819 & 1.0731 \\
lsc-doma-1m0-05-kb78 & V & 2456343.68000 & 21.011 & 0.026 & 38.004 & 3.224 & 1.451 & 0.976 & 1.0451 \\
\vdots   & \vdots & \vdots  & \vdots & \vdots & \vdots & \vdots   & \vdots  & \vdots & \vdots \\
lsc-domb-1m0-09-kb73 & I & 2456364.58433 & 20.885 & 0.019 & 43.815 & 2.107 & 0.426 & 0.787 & 1.0018 \\
lsc-domb-1m0-09-kb73 & I & 2456364.58801 & 20.923 & 0.020 & 43.815 & 2.107 & $-$1.077 & 0.786 & 1.0024 \\
\hline
\end{tabular}
\label{tab:vri_phot}
\end{table*}

\end{document}